\documentclass[11pt,a4paper,final]{iopart}

\expandafter\let\csname equation*\endcsname\relax

\expandafter\let\csname endequation*\endcsname\relax

\usepackage{iopams}  
\usepackage{graphicx}
\usepackage[breaklinks=true,colorlinks=true,linkcolor=blue,urlcolor=blue,citecolor=blue]{hyperref}
\usepackage[english]{babel}
\usepackage{siunitx}
\usepackage{graphicx}
\usepackage{mathtools}
\usepackage{setspace}
\usepackage{verbatim}
\usepackage{caption}
\usepackage{wrapfig}
\usepackage{color}

\usepackage{booktabs}
\usepackage[toc,page]{appendix}

\usepackage[doi=false,isbn=false,url=false,eprint=false,
    backend=biber, 
    natbib=true,
    style=numeric,
    sorting=none,
]{biblatex}
\begin{document}

\title[]{\vspace{-3.5cm}\\Two sides of the same coin: the $\mathcal{F}$-statistic and the 5-vector method}

\author{ L. D'Onofrio$^1$, P. Astone$^{1}$, S. Dal Pra$^{1}$, S. D'Antonio$^{^1}$, 
M. Di Giovanni$^{2,1}$, R. De Rosa$^{3,4}$, P. Leaci$^{2,1}$, S. Mastrogiovanni$^1$, L. Mirasola$^{5,6}$, F. Muciaccia$^{2,1}$, C. Palomba$^1$, L. Pierini$^{1}$ }
\address{$^1$INFN, Sezione di Roma, I-00185 Roma, Italy
\\$^2$Universit\`a di Roma La Sapienza, 00185 Roma, Italy
\\$^3$Università di Napoli “Federico II”, I-80126 Napoli, Italy
\\$^4$INFN, Sezione di Napoli, I-80126 Napoli, Italy
\\$^5$Universit\`a degli Studi di Cagliari, Cagliari  09042, Italy
\\$^6$INFN, Sezione di Cagliari, Cagliari 09042, Italy}

\ead{luca.donofrio@roma1.infn.it}

\begin{abstract}
This work explores the relationship between two data-analysis methods used in the search for continuous gravitational waves in LIGO-Virgo-KAGRA data: the $\mathcal{F}$-statistic and the 5-vector method. We show that the 5-vector method can be derived from a maximum likelihood framework similar to the $\mathcal{F}$-statistic. Our analysis demonstrates that the two methods are statistically equivalent, providing the same detection probability for a given false alarm rate. We extend this comparison to multiple detectors, highlighting differences from the standard approach that simply combines 5-vectors from each detector. In our maximum likelihood approach, each 5-vector is weighted by the observation time and sensitivity of its respective detector, resulting in efficient estimators and analytical distributions for the detection statistic. Additionally, we present the analytical computation of sensitivity for different searches, expressed in terms of the minimum detectable amplitude.
\end{abstract}

\section{Introduction}
Spinning neutron stars with a non-axisymmetric mass distribution are promising targets for the emission of continuous gravitational wave (CW) radiation in the LIGO-Virgo-KAGRA \cite{ligo,virgo,kagra} (LVK) observational frequency band. 

At the source, the CW signal is quasi-monochromatic with the gravitational wave (GW) frequency $f_{gw}$ that is proportional to the source rotation frequency according to the considered emission model \cite{cwemission}. At the detector, the CW signal has a phase modulation mainly due to the Doppler effect produced by the relative motion between the source and the Earth. Due to the Earth sidereal motion and the response of the detector to the coming signal, there is also an amplitude and phase modulation that splits the signals in five frequencies $f_{gw}\,,f_{gw}\pm f_\oplus\,,f_{gw}\pm 2 f_\oplus$, where $f_\oplus$ is the Earth sidereal frequency \cite{2010}.

So far, there is no significant evidence of a CW signal in LVK data (for reference, see \cite{ornella}). Latest results from the most sensitive analysis targeting known pulsars - the so-called \textit{targeted searches} - set interesting upper limits \cite{mio3,O3targ} that are now approaching theoretical constraints \cite{cwemission} on the pulsar ellipticity, i.e. the physical parameter that quantifies the mass distribution asymmetry with respect to the rotation axis.

In the context  of the search for CWs in LVK data, two of the most sensitive pipelines developed in the last decades are the $\mathcal{F}$-statistic \cite{Fstat} and the 5-vector method \cite{2010}. 
Both methods are frequentist pipelines, generally defined as matched filtering techniques applied in the time-domain and in the frequency-domain, respectively. Although this definition is intuitive, it is theoretically not accurate since the matched filtering theory requires the exact knowledge for the expected signal. Indeed, even though the CW signal has a clear signature at the detector, the exact shape of the signal depends on two unknown parameters that fix the GW polarizations.

The $\mathcal{F}$-statistic is inferred \cite{JKS} maximizing the likelihood ratio on the unknown parameters leaving the dependence on the source parameters that can be accurately known, as in targeted searches, or generally fixed from an optimized grid. Originally, the 5-vector method \cite{2010} has not been inferred from a maximum likelihood principle. 
Based on a complex-number formalism, the expected signal can be written in terms of two (plus and cross) polarization amplitudes. 
The 5-vector method defines two matched filters in the frequency domain for both the plus and cross polarization. The associated detection statistic corresponds to the linear combination of the squared modulus of these two matched filters \cite{2010}. The multidetector extension in \cite{2014} combines the 5-vectors from each detector defining the so-called 5n-vector, where $n$ is the number of considered detectors. 

In this paper, we show that the 5-vector method statistic can be inferred from a maximum likelihood approach with a re-definition of \textit{ad hoc} coefficients. The noise and signal distributions of the inferred statistic are equivalent to the  $\mathcal{F}$-statistic distributions showing that the two methods are statistically equivalent. We generalize the procedure to $n$ detectors showing important differences with respect to the standard 5n-vector definition in \cite{2014}. 

The paper is organized as follows. In Section \ref{Sec:ML}, we provide an overview of maximum likelihood statistics introducing the $\mathcal{F}$-statistic. In Section \ref{Sec:5vec}, we review the formalism and the state-of-art of the 5-vector method. In Section \ref{Sec:5vecML}, we infer the 5-vector statistic from a maximum likelihood approach generalizing to the multidetector case. In Section \ref{Sec:app}, we describe some applications of the maximum likelihood statistics including the theoretical estimation of the minimum detectable signal amplitude for different types of searches.

\section{Maximum likelihood approach}\label{Sec:ML}
Assuming a signal $h(t)$ and additive noise $n(t)$, the detector output $x(t)$ can be written as:
\begin{equation}
    x(t)=n(t)+h(t) \,.
\end{equation}
The likelihood ratio~\cite{JKS} is defined as the probability to have a signal in the analyzed data divided by the probability to have just noise. It can be shown that for Gaussian noise: 
\begin{equation}
    \ln{\Lambda}\equiv \ln{\frac{P(x|h)}{P(x|h=0)}}=\left( x|h \right) - \frac{1}{2} \left( h|h \right),
\end{equation} 
where the scalar product is defined for a small frequency band as:
\begin{equation}\label{scalarprod}
    \left( a|b \right)= 2 \int_f^{f+\delta f} \frac{\Tilde{a}(f')\Tilde{b}(f')}{S_h(f')}df'\cong\frac{2}{S_h} \int_0^{T_\textit{obs}} a(t)b^*(t) dt, 
\end{equation}
with $S_h$ being the single-sided power spectral density, which we assume constant in a narrow frequency band around the expected CW frequency. The likelihood depends on unknown parameters and, for composite hypothesis, the Neymann-Pearson criteria \cite{NPlemma} can not be applied. A simple and common approach to construct a detection statistic in a frequentist framework is to consider the likelihood ratio maximized over the unknown parameters. 

As shown in \cite{notopt}, statistics inferred from the estimation of the maximum likelihood (hereafter MLE statistics) are not ``optimal'' in the Neyman-Pearson sense since Bayesian methods can be more powerful considering priors consistent with the parameter distribution.

\subsection{$\mathcal{F}$-statistic} \label{sec:fstat}
The $\mathcal{F}$-statistic  is  obtained  maximizing  the likelihood function with respect to the four signal parameters: the signal amplitude $h_0$, the inclination angle $\iota$ of the neutron star rotational axis with respect to the line of sight, the wave polarization $\psi$ and the initial phase $\phi_0$, while keeping the dependence on the source parameters (sky position and rotational parameters). 

The likelihood can be expressed more clearly by rewriting the expected signal as a linear combination of four basis terms~\cite{JKS} each with amplitude $\lambda^a$ : 
\begin{equation}\label{expsigF}
    h(t)=\sum_{a=1}^4 \lambda^a h_a(t)\,.
\end{equation}Each term $h_a(t)$ corresponds to a particular combination of the phase evolution $\Phi(t)$ and of the sidereal modulation, which depends on the antenna pattern that defines the response of the detector to the GW signal. Then, the $\mathcal{F}$-statistic, defined to be twice the logarithm of the maximized likelihood ratio, is:
\begin{equation}\label{fstatdef}
\mathcal{F}=(x|x)-(x- \sum_{a=1}^4 \lambda^a h_a(t)|x-\sum_{b=1}^4 \lambda^b h_b(t) )=\sum_{a,b=1}^4 (\Gamma^{-1})^{a,b}  (x|h_a)(x|h_b)
\end{equation}where the matrix $\Gamma$ is the Fisher matrix \cite{genfstat}, defined as:
\begin{equation}
    \Gamma^{a,b}\equiv \left( \frac{\partial h}{\partial \lambda_a}\bigg\rvert\frac{\partial h}{\partial \lambda_b} \right)=(h_a|h_b) \,,
\end{equation}and the maximized values for $\lambda_a$ are:
\begin{equation}
    \overline{\lambda}_a=\sum_{b=1}^4 (\Gamma^{-1})^{a,b}  (x|h_b)
\end{equation}

Assuming stationary Gaussian noise, it is easy to show \cite{Fstat} that the $\mathcal{F}$-statistic satisfies a  $\chi^2$ distribution with 4 degrees of freedom  and, in the presence of a signal, it has a non-centrality parameter equal to the squared optimal signal-to-noise ratio (SNR) \cite{metricF}: $\rho^2\equiv (h(t)|h(t))$.

\section{The 5-vector method}\label{Sec:5vec}
The difference between the $\mathcal{F}$-statistic and the 5-vector method arises from the formalism used for the expected signal $h(t)$. In the 5-vector context\footnote{Assuming that we have heterodyned data, corrected for the Doppler and spin-down modulation (see \cite{2019} for more details).}, the signal is written as the real part of:
\begin{equation}\label{5vech}
    h(t)=H_0\textbf{A}\cdot \textbf{W}e^{j\omega_0 t}\equiv H_0 \left( H_+\textbf{A}^+ + H_\times\textbf{A}^\times\right)\cdot \textbf{W} e^{j\omega_0 t}  \,.
\end{equation}
In bold, we will refer to an array of five complex components and the product $\mathbf{B}\cdot \mathbf{C}$ is defined as $\mathbf{B}\cdot \mathbf{C}=\sum_{i=1}^5 B_i C_i^*$. The vector $\textbf{W}$ in \eqref{5vech} is $\textbf{W}=e^{jk\Theta}$ with $k=0,\pm1,\pm2$ and $\Theta$, the local sidereal angle \cite{2010}\footnote{As in \cite{2010}, we will indicate with $\textbf{W}^*\equiv e^{-ik\Omega_\oplus t}=\textbf{W}e^{ik(\alpha-\beta)}$ the \textit{5-vector generator} where $\alpha$ is  the source right ascension and $\beta$ the detector longitude.}. 
\\The amplitude $H_0$ is linked to the classical amplitude $h_0$ by:
\begin{equation}\label{corramp}
 H_0=h_0 \sqrt{\frac{1+6\cos^2\iota + \cos^4 \iota}{4}}\,,
\end{equation}
while $H_{+/\times}$ are the polarization functions, 
\begin{equation}
H_+=\frac{\cos(2\psi)-i\eta \sin(2\psi)}{\sqrt{1+\eta^2}} \qquad    H_\times=\frac{\sin(2\psi)+i\eta \cos(2\psi)}{\sqrt{1+\eta^2}}
\end{equation}
that depend on the polarization parameters $\psi$ and $\eta$:
\begin{equation}
\eta=-\frac{2\cos \iota}{1+\cos^2 \iota}\,.
\end{equation} 
The extended expressions of the 5 components of the so-called 5-vector template $\textbf{A}^{+/\times}$, which include the detector response to the GW signal, can be found in \cite{2010}. 

The signal in \eqref{5vech} is composed of two templates that depend on two overall polarization amplitudes, $H_0H_{+/\times}$. For each template, it can be applied a matched filter that corresponds to a maximum likelihood approach to estimate each amplitude\footnote{For more details, see \cite{magg} Section 7.4.2}. 
\\The statistic is then inferred from the estimators (i.e. the matched filters) of the plus and cross polarization amplitudes $H_0H_{+/\times}$:
\begin{equation}\label{estim}
    \hat{H}_{+/\times}=\frac{\textbf{X} \cdot \textbf{A}^{+/\times}}{|\textbf{A}^{+/\times}|^2}\,,
\end{equation}where 
\begin{equation}\label{5vec}
 \textbf{X} =\int_{T_{obs}}x(t) e^{-j\textbf{k} \Omega_\oplus t} e^{-j\omega_0t}dt \qquad \text{with} \qquad \textbf{k}=(0,\pm 1,\pm 2),
\end{equation} is the data 5-vector computed from the detector data, $x(t)$. In \cite{2010}, the statistic is defined as:
\begin{equation}
    S=|\textbf{A}^+|^4 |\hat{H}_+|^2  + |\textbf{A}^\times|^4 |\hat{H}_\times|^2
\end{equation}that improves the ROC curve with respect to considering equal coefficients or just squared coefficients  $|\textbf{A}^+|^2$ if the two polarization have different "weights" (see Figure 1 in \cite{2010}).

To generalize the statistic to the multidetector case, the estimators in \eqref{estim} are defined using the 5n-vector \cite{5n}, i.e. the combination of the 5-vector from each of the $n$ considered detectors: $\mathbf{X}=[\mathbf{ X}_1,..., \mathbf{X}_n]$ and $\mathbf{A}^{+/\times}=[\mathbf{A}_1^{+/\times},..., \mathbf{A}_n^{+/\times}]$. 

In \cite{mio3}, it is defined a normalized statistic $\tilde S$ for the multidetector analysis:
\begin{equation}\label{statS}
\tilde S= \frac{|\textbf{A}^{+}|^4 }{\sum_{j=1}^n \sigma_j^2 \, T_{j} \, |\textbf{A}^{+}_{j}|^2} |\hat{H}_+|^2 +  \frac{|\textbf{A}^{\times}|^4 }{\sum_{k=1}^n \sigma_k^2 \, T_{k} \, |\textbf{A}^{\times}_{k}|^2} |\hat{H}_\times|^2 \,.
\end{equation}
where $\sigma_j^2$ and $T_j$ are the variance of the data distribution in a frequency band around $f_{gw}$ (usually few tenths of Hz wide) and the observation time in the $j^{\text{th}}$ detector, respectively. In the case of Gaussian noise, the distribution of the statistic $\tilde S$ in Eq.~\eqref{statS} is a Gamma distribution $\Gamma(x;2,1)$ with shape $2$ and scale parameter $1$.
\\If a signal is present, the distribution of $\tilde S$ is proportional to a 4-D $\chi^2$ distribution with non centrality parameter $\lambda$:
\begin{equation}
    \tilde S \sim 2\cdot \chi^2(2x;4,\lambda)\,,
\end{equation}
\begin{equation}\label{Lambda}
    \lambda=2 H_{0}^2 \left(\frac{|\textbf{A}^{+}|^4 |H_{+}|^2}{\sum_{j=1}^n \sigma_j^2 T_{j} \, |\textbf{A}^{+}_{j}|^2}+\frac{|\textbf{A}^{\times}|^4 |H_{\times}|^2}{\sum_{k=1}^n \sigma_k^2 T_{k} \,|\textbf{A}^{\times}_{k}|^2}\right)\,.
\end{equation}
For more details on $\tilde S$, we refer to the work in \cite{mio2}.

\section{Maximum likelihood approach to the 5-vector}\label{Sec:5vecML}\label{Sec1}
In this Section, we infer the 5-vector method from a maximum likelihood approach. First, we consider the single detector case and then, we generalize to the multidetector case considering different observation time and sensitivity for each detector.

\subsection{Single detector}
Following \cite{JKS}, we can express the likelihood as
\begin{equation}
        \ln{\Lambda}=\textit{Re}\left\{\left( x|h_+ \right) - \frac{1}{2} \left( h_+|h_+ \right)+\left( x|h_\times \right) - \frac{1}{2} \left( h_\times|h_\times \right) \right\}
\end{equation} where
\begin{equation}
    h_{+/\times}=H_0H_{+/\times}\textbf{A}^{+/\times}\cdot \textbf{W} e^{j\omega_0 t} \,.
\end{equation} Indeed, due to the orthogonality of the GW polarization, it follows that $\left(h_+| h_\times\right)=\left(h_\times| h_+\right)=0$.

The polarization amplitudes $H_0H_{+/\times}$ correspond to the $\lambda^\alpha$ in \eqref{expsigF} with the difference that in the 5-vector formalism, we consider the linear combination of only two terms due to their complex nature.
Evaluating the scalar product defined in \eqref{scalarprod}
\begin{equation}
    (x|h_+)=\frac{2T_0}{S_h} \left(\frac{1}{T_0} \int_0^{T_0} x(t)\textbf{W}^* e^{-j\omega_0 t} dt \right) H_0H_+^*(\textbf{A}^+)^* =\frac{2T_0}{S_h}H_0H_+^*\textbf{X}\cdot\textbf{A}^+ \,,
\end{equation}
we can introduce the data 5-vector $\textbf{X}$\footnote{Note that the 5-vector definition has here an additional factor $T_0^{-1}$ with respect to \cite{2010} that is important for the multidetector extension.}, i.e. the Fourier components of the data at the expected five frequencies:
\begin{equation}\label{5vecnew}
    \textbf{X}= \frac{1}{T_0} \int_0^{T_0} x(t)\textbf{W}^* e^{-j\omega_0 t} dt \,,
\end{equation}
and also the product between 5-vectors $\textbf{A}\cdot\textbf{B}=\sum A_i B_i^*$, in analogy with \cite{2010}. The second term in the likelihood is\footnote{It should be noted that we have defined $|\textbf{A}^+|^2\equiv \textbf{A}^+ \cdot \textbf{A}^+$ in terms of 5-vector product.}:
\begin{equation}
(h_+|h_+)=\frac{2}{S_h} H_0^2|H_+|^2|\textbf{A}^+|^2\left(\int_0^{T_0} dt \right) = \frac{2T_0}{S_h} H_0^2|H_+|^2|\textbf{A}^+|^2 \,.
\end{equation}
It follows that:
\begin{equation}
    \ln\Lambda=\frac{T_0}{S_h} \sum_p^{+,\times}\left[H_0H_p^*\textbf{X}\cdot\textbf{A}^p + H_0H_p(\textbf{X}\cdot\textbf{A}^p)^*-H_0^2|H_p|^2|\textbf{A}^p|^2\right] 
\end{equation}
The derivatives with respect to the complex polarization amplitudes are:
\begin{equation}
    \frac{\partial \ln \Lambda}{\partial H_0H_{+/\times}^*}\propto  (\textbf{X}\cdot\textbf{A}^{+/\times}) - H_0H_{+/\times} |\textbf{A}^{+/\times}|^2
\end{equation}that are set to zero if:
\begin{equation}
    \left(H_0H_{+/\times}\right)_{\text{MAX}} \equiv \hat{H}_{+/\times}=\frac{\textbf{X} \cdot \textbf{A}^{+/\times}}{|\textbf{A}^{+/\times}|^2} \,.
\end{equation}
The maximized likelihood ratio with these estimators results in:
\begin{equation}\label{SasF}
    (\ln \Lambda)_{\text{MAX}} = \frac{T_0}{S_h} \left( |\textbf{A}^+|^2 |\hat{H}_+|^2  + |\textbf{A}^\times|^2 |\hat{H}_\times|^2 \right)\varpropto |\textbf{A}^+|^2 |\hat{H}_+|^2  + |\textbf{A}^\times|^2 |\hat{H}_\times|^2 \,.
\end{equation}The common factor $T_0/S_h$ is irrelevant since it does not influence the ROC curves. The results is in agreement with \cite{2010}, where it is stated: \textit{"If we take the two coefficients
proportional to the square of the absolute values of the signal 5-vectors (i.e. $\textbf{A}^{+/\times}$), we have the well-known F-statistics, which is an equalization of the response at the two
modes"}. The same result can be also found in \cite{phasedecomp} but expressed in the $\mathcal{F}$-statistic formalism. 

It easy to show that the noise and signal distributions of $2(\ln \Lambda)_{\text{MAX}}$ correspond to the $\mathcal{F}$-statistic distributions showing that the two methods are statistically equivalent, i.e. for a fixed false alarm probability they provide the same detection probability.

The definition of the data 5-vector in \eqref{5vecnew} with the factor $1/T_0$ entails the constant factor $T_0/S_h$ in \eqref{SasF}. Considering the "normalized" definition in \eqref{statS} with $n=1$, the constant factor is $1/(T_0S_h)$; the difference arises due to the different definition of the data 5-vector. The weighted definition of the data 5-vector is important for the multidetector extension, as shown in the next Section. 


\subsection{Multidetector}
Let us consider $n$ detectors assuming stationary Gaussian noise with different variance and observation time for each detector. The new form of the likelihood depends on the expression of the scalar product in \eqref{scalarprod} that is (assuming uncorrelated noise):
\begin{equation}
    \left( \textbf{a}|\textbf{b} \right)\cong 2 \sum\limits_{i=1}^n  \int_0^{T_i} \frac{a_i(t)b_i^*(t)}{S_i} dt \,.
\end{equation}The scalar products in the multidetector case are:
\begin{equation}
    \left( \textbf{x}|\textbf{h}_{+/\times} \right)= 2 H_0H_{+/\times}^* \sum\limits_{i=1}^n \frac{T_i}{S_i}\left(\textbf{X}_i\cdot \textbf{A}^{+/\times}_i\right)
\end{equation}where $\textbf{X}_i$ is the data 5-vector for the i-th detector, as defined in \eqref{5vecnew}. Since
\begin{equation}
    \left( \textbf{h}_{+/\times}|\textbf{h}_{+/\times} \right) = 2 H_0^2 |H_{+/\times}|^2 \sum\limits_{i=1}^n \frac{T_i|\textbf{A}^{+/\times}_i|^2}{S_i} \,,
\end{equation}the likelihood is maximized for:
\begin{equation}\label{estmax}
\left(H_0H_{+/\times}\right)_{\text{MAX}}\equiv \hat{H}_{+/\times} = \left( \sum\limits_{j=1}^{n} \frac{T_j\,\textbf{X}_j \cdot \textbf{A}_j^{+/\times}}{S_j}\right) \left( \sum\limits_{k=1}^{n}  \frac{ T_k|\textbf{A}_k^{+/\times}|^2}{S_{k}}\right)^{-1}\,.
\end{equation}
This is quite different from the standard definition of the 5n-vector in \cite{5n} where the estimators are generalized from the single detector case:
\begin{equation}\label{estim5n}
\hat{H}_{+/\times} =\frac{\textbf{X} \cdot \textbf{A}^{+/\times}}{|\textbf{A}^{+/\times}|^2}=\left(\sum\limits_{j=1}^{n}\textbf{X}_j \cdot \textbf{A}_j^{+/\times}\right)\left(\sum\limits_{k=1}^{n}|\textbf{A}_k^{+/\times}|^2\right)^{-1}\,,
\end{equation}
where $\textbf{X}=[\textbf{X}_1,...,\textbf{X}_n]$ and $\textbf{A}^{+/\times}=[\textbf{A}^{+/\times}_1,...,\textbf{A}^{+/\times}_n]$.
The two estimators definition in \eqref{estmax} and \eqref{estim5n} coincide if each detector has the same observation time and sensitivity.

From the maximization of the likelihood, it follows that:
\begin{equation}\label{lambda+5n}
    (\ln \Lambda)_{\text{MAX}} =\sum\limits_{p=+,\times} \left(\sum\limits_{k=1}^n \frac{T_k|\textbf{A}^p_k|^2}{S_k}\right)^{-1} \left[ \left( \sum\limits_{i=1}^n \frac{T_i\,\textbf{X}_i\cdot\textbf{A}^p_i}{S_i}  \right)\left( \sum\limits_{j=1}^n \frac{T_j\,\textbf{X}_j\cdot\textbf{A}^p_j}{S_j}  \right)^*\right] \,.
\end{equation}
From \eqref{lambda+5n}, if we define the weighted data 5n-vectors as
\begin{equation}\label{5nvec_ML}
\Tilde{\textbf{X}}=\left[\sqrt{\frac{T_1}{S_1}}\, \textbf{X}_1,...,\sqrt{\frac{T_n}{S_n}}\, \textbf{X}_n\right] 
\end{equation}
and the weighted template 5n-vectors
\begin{equation}
    \Tilde{\textbf{A}}^{+/\times}=\left[\sqrt{\frac{T_1}{S_1}}\, \textbf{A}^{+/\times}_1,...,\sqrt{\frac{T_n}{S_n}}\, \textbf{A}^{+/\times}_n\right]\,,
\end{equation}
we can re-write the maximum likelihood statistics:
\begin{equation}\label{lambdamax}
    (\ln \Lambda)_{\text{MAX}}=|\Tilde{\textbf{A}}^+|^2 |\Tilde{H}_+|^2 + |\Tilde{\textbf{A}}^\times|^2 |\Tilde{H}_\times|^2
\end{equation}
where 
\begin{equation}
    \Tilde{H}_{+/\times}= \frac{\textbf{X} \cdot \Tilde{\textbf{A}}^{+/\times}}{|\Tilde{\textbf{A}}_k^{+/\times}|^2} \equiv\left(\sum\limits_{j=1}^{n}\Tilde{\textbf{X}}_j \cdot \Tilde{\textbf{A}}_j^{+/\times} \right)  \left(\sum\limits_{k=1}^{n}|\Tilde{\textbf{A}}_k^{+/\times}|^2\right)^{-1}\,.
\end{equation}The statistic \eqref{lambdamax} has the same expression of the $\mathcal{F}$-statistic that we have seen before, but with weighted 5n-vectors. To explore the relation between \eqref{lambdamax} and \eqref{statS}, let us consider the noise distribution of these statistics. From \cite{mio3}, $\Tilde S\sim \Gamma(x;2,1)$ that is proportional to a 4-D $\chi^2$ distribution (hence, it is the $\mathcal{F}$-statistic).  To infer the distribution of $(\ln \Lambda)_{\text{MAX}}$, we start from:
\begin{equation}
    \Tilde{H}_{+/\times} \sim Gauss(x;0,\Tilde{\sigma}^2_{+/\times}) \qquad |\Tilde{H}_{+/\times}|^2\sim Exp(x; \Tilde{\sigma}^2_{+/\times})
    \end{equation}with
    \begin{equation}\label{sigmatilde}
    \Tilde{\sigma}^2_{+/\times}=\sum\limits_{j=1}^{n}\frac{T_j(S_j)^{-1} S_j T_j^{-1} |\Tilde{\textbf{A}}_j^{+/\times}|^2}{|\Tilde{\textbf{A}}^{+/\times}|^4}=\frac{1}{|\Tilde{\textbf{A}}^{+/\times}|^2}\,.
\end{equation}The factor $T_j(S_j)^{-1}$ comes from the weight of the data 5-vector, while $S_j T_j^{-1}$ from the definition of the 5-vector in \eqref{5vecnew}. The coefficients in \eqref{lambdamax} are exactly $1/\Tilde{\sigma}^2_{+/\times}$ that normalize the distribution to the $\Gamma(x;2,1)$ as in the case of $\tilde S$ in \eqref{statS}.

For $n$ co-located detectors, $\,|\textbf{A}_k^{+/\times}|^2=|\textbf{A}^{+/\times}_0|^2 \,,\forall \, k$ with the same observation time $T_0$, it follows that:
\begin{equation}
     \Tilde{\sigma}^2_{+/\times}=\left( \sum\limits_{k=1}^{n} \frac{T_k |\textbf{A}_k^{+/\times}|^2}{S_k} \right)^{-1}=\frac{1}{T_0|\textbf{A}^{+/\times}_0|^2}\left( \sum\limits_{k=1}^{n} \frac{1}{S_k} \right)^{-1}= \frac{1}{T_0|\textbf{A}^{+/\times}_0|^2} \frac{\mathcal{H}}{n}\,,
\end{equation}where $\mathcal{H}$ is the harmonic mean of the $S_j$, i.e. $n$ co-located detectors corresponds to a single detector with sensitivity equal to the harmonic mean divided by $n$. The relation
\begin{equation}
min\{S_1,...,S_n\}\leq \mathcal{H} \leq  n\, min\{S_1,...,S_n\}
\end{equation}entails that:
\begin{equation}\label{imp}
\frac{min\{S_1,...,S_n\}}{n}\leq \frac{\mathcal{H}}{n} \leq  min\{S_1,...,S_n\} \,.
\end{equation}
For $n$ co-located detectors with the same observation time, we always have an improvement in the detection sensitivity, differently from what is found for the classic definition\footnote{For more details, see Appendix A in \cite{mio3}.} of the 5n-vector

The difference between \eqref{lambdamax} and \eqref{statS} is in the estimators \eqref{estmax}. The parameter estimation improves (i.e. has smaller variance) using these estimators that maximize the likelihood, as analyzed in Section \ref{subsec:Fisher}. 

\subsection{Real data}
In real analysis, we cannot use the theoretical expressions in \cite{2010} for $\textbf{A}^{+/\times}$ as the matched filters would not take into
account the presence of gaps in the data and the effect of all pre-processing operations. 

In \cite{2010}, it is proposed to approximate the signal 5-vector simulating in the time domain the
signal $+/\times$ components with the theoretical expressions in \cite{2010}, but considering gaps and all the pre-processing procedures, obtaining two time series $s^{+/\times}(t)$. The signal 5-vector are hence defined as:
\begin{equation}
    \textbf{A}^{+/\times}=\frac{1}{T_{0}}\int_0^{T_0} s^{+/\times}(t)\textbf{W}^* e^{j\omega_0 t} dt\,,
\end{equation}
i.e. computing the Fourier transform\footnote{Given the finite duration of the Fourier transform, part of the energy of the Fourier components will be spread into lateral bands decreasing the power of the signal.} at the expected five frequencies as in the case of the data 5-vector. The results shown so far are not influenced by this definition using real data since $\textbf{A}^{+/\times}$ have a constant value fixing the source and detector position.

\section{Application}\label{Sec:app}
In this Section, we briefly summarize some useful applications and properties of the MLE 5-vector. First, considering the Fisher matrix formalism, we show that the estimators in \eqref{estmax} are statistically efficient. Then, we set a solid framework for the construction of a grid in the parameter space. In the last Section, we provide a theoretical estimation of the minimum detectable amplitude for different searches.

\subsection{Fisher matrix}\label{subsec:Fisher}
The $\mathcal{F}$-statistic is usually written in terms of the Fisher matrix as shown in Section \eqref{sec:fstat}. In the 5-vector formalism, the elements of the Fisher matrix are:
\begin{equation}
    \Gamma^{++}=\left( \frac{\partial \textbf{h}(t)}{\partial H_0H_+}\bigg\vert\frac{\partial \textbf{h}(t)}{\partial H_0H_+} \right)=\sum\limits_{k=1}^{n} \frac{T_k|\textbf{A}_k^{+}|^2}{S_k}\,,
\end{equation}
\begin{equation}
    \Gamma^{\times\times}=\left( \frac{\partial \textbf{h}(t)}{\partial H_0H_\times}\bigg\vert\frac{\partial \textbf{h}(t)}{\partial H_0H_\times} \right)=\sum\limits_{k=1}^{n} \frac{T_k|\textbf{A}_k^{\times}|^2}{S_k}\,,
\end{equation}
\begin{equation}
    \Gamma^{+\times}=\left( \frac{\partial \textbf{h}(t)}{\partial H_0H_+}\bigg\vert\frac{\partial \textbf{h}(t)}{\partial H_0H_\times} \right)=\sum\limits_{k=1}^{n} \frac{T_k\textbf{A}_k^{+}\cdot (\textbf{A}_k^{\times})^*}{S_k}=0=(\Gamma^{\times +})^*\,.
\end{equation}The $\Gamma^{+\times}$ (and hence $\Gamma^{\times +}$) goes to zero since $\textbf{A}_k^{+}\cdot (\textbf{A}_k^{\times})^*=0 \,, \forall k$ from the theoretical expressions in \cite{2010} due to the orthogonality of the GW polarizations.
\\The diagonal elements are the inverse of the coefficients in \eqref{lambda+5n} since in the Fisher formalism, the MLE statistic is expressed as: 
\begin{equation}
(\ln \Lambda)_{\text{MAX}}=\sum_{p}^{+/\times} (\Gamma^{-1})^{pp}  (\textbf{x}|\textbf{h}_p)(\textbf{x}|\textbf{h}_p)\,.
\end{equation}
The maximum likelihood estimators can be also written in terms of the Fisher matrix:
\begin{equation}
    \hat{H}_+ \equiv (\Gamma^{-1})^{++}\left(\textbf{x}\bigg\vert\frac{\partial \textbf{h}(t)}{\partial H_0H_+}\right)\,,
\qquad
    \hat{H}_\times \equiv (\Gamma^{-1})^{\times\times}\left(\textbf{x}|\frac{\partial \textbf{h}(t)}{\partial H_0H_\times}\right)\,,
\end{equation}
and the inverse of the Fisher matrix element is a lower bound (the so-called Cramér-Rao bound) on the variance of any unbiased estimator:
\begin{equation}
    (\Gamma^{-1})^{++/\times\times}\equiv \Tilde{\sigma}^2_{+/\times}   
\end{equation}with $\Tilde{\sigma}^2_{+/\times}$ defined in \eqref{sigmatilde}. Indeed, the Fisher matrix is often called
the covariance matrix, since the
ML estimators distribution tends
to a jointly Gaussian distribution with covariance matrix equal to $\Gamma^{-1}$.
\\The variances $\sigma^2_{+/\times}$ for the estimators defined in \eqref{estim5n} are:
\begin{equation}
\sigma^2_{+/\times}=\sum\limits_{j=1}^{n}\frac{ S_j\cdot T_j \cdot |\textbf{A}_j^{+/\times}|^2}{|\textbf{A}^{+/\times}|^4}\,,
\end{equation}
and the Cramér-Rao bound entails that $\sigma^2_{+/\times}>\tilde \sigma^2_{+/\times}$, i.e. the variance of the estimators in \eqref{estmax} is smaller with respect to the variance of the estimators with the standard definition in \eqref{estim5n}. 

Since the signal distributions of the estimators in \eqref{estmax} are:
\begin{equation}
    \Tilde{H}_{+/\times} \sim Gauss\left(x;\frac{2H_0H_{+/\times}}{\Tilde{\sigma}^2_{+/\times}},\Tilde{\sigma}^2_{+/\times}\right) \,,
    \end{equation}
the signal distribution of $S$ is proportional to a non-central $\chi^2$ distribution,
\begin{equation}
(\ln \Lambda)_{\text{MAX}}\sim 2\cdot \chi^2(2x;4,\rho^2)\,,
\end{equation}
where the parameter $\rho^2 $ is:
\begin{equation}\label{lambdanew}
    \rho^2 \equiv (\textbf{h}_+|\textbf{h}_+)+(\textbf{h}_\times|\textbf{h}_\times) = H_0^2\sum_p^{+/\times} \left( \frac{2|H_p|^2}{\Tilde{\sigma}^2_p} \right)\,,
\end{equation}i.e. the optimal signal-to-noise ratio. For the Cramér-Rao bound, $\rho^2$ is the largest non-centrality parameter fixing the source parameters. It follows that the new definition of the 5n-vector in \eqref{5vecnew} maximize the optimal signal-to-noise ratio.

\subsection{Phase metric}
If either the source parameters are not known with the required accuracy for a targeted search or when performing an \textit{all-sky} search, a template grid is usually built in the search parameter space. In the last years, several works (see for example \cite{metricF,allskyF,binaryF}) have described the implementation of a template grid using the $\mathcal{F}$-statistic. Generally,  the grid is chosen minimizing the distance between any point in parameter space and the
nearest grid point. Recently, \cite{optimalmetric} showed that standard template banks do not always maximize the expected number of detections and for high dimensional space parameters (above eight dimensions), random template banks outperform the best known lattices~\cite{randommetric} .  

A first attempt to set a template grid using the 5-vector formalism is described in \cite{phasedecomp}, where the authors refer to the statistic \eqref{statS} with squared coefficients  as "$\mathcal{F}$-statistic". In this Section, we generalize the results in \cite{phasedecomp} to the multidetector case using a slightly different approach and following \cite{metricF}. 

The signal distribution of the MLE statistic in \eqref{lambdamax} is a 4-D $\chi^2$ distribution with non centrality parameter $\rho^2$.
We can generalize the results to the multidetector case evaluating the non centrality parameter, expliciting \eqref{lambdanew}:
\begin{equation}
    \rho^2 = 2 H_0^2 \left(|H_+|^2 \sum\limits_{i=1}^n \frac{T_i|\textbf{A}^+_i|^2}{S_i} + |H_\times|^2 \sum\limits_{i=1}^n \frac{T_i|\textbf{A}^\times_i|^2}{S_i}  \right) \,.
\end{equation}
This optimal (i.e. from the optimal filter) SNR does not depend on the initial phase, and $\rho$ scales linearly with the overall amplitude and with the square root of the observation time, as expected.

Let us suppose that we are searching with a template set parameters $\Vec{\theta}_t$ a CW signal with real parameters $\Vec{\theta}$, where $\Vec{\theta}_t=\Vec{\theta}+\Delta\Vec{\theta}$. The mismatch function is defined as:
\begin{equation}
    \mu(\mathcal{A},\Vec{\theta}_t;\Vec{\theta})=\frac{\rho^2(\mathcal{A},\Vec{\theta})-\rho^2(\mathcal{A},\Vec{\theta}_t)}{\rho^2(\mathcal{A},\Vec{\theta})}\,,
\end{equation}and it can be written as (assuming summation on repeated indexes):
\begin{equation}
\mu(\mathcal{A},\Vec{\theta}_t;\Vec{\theta})=g_{ij}(\mathcal{A},\Vec{\theta}_t)\Delta\theta_i\Delta\theta_j + \mathcal{O}(\Delta\Vec{\theta}^3)
\end{equation}where $g_{ij}$ is the normalized projected Fisher matrix as defined in \cite{metricF}, depending on the unknown polarization amplitudes $\mathcal{A}$.  A more
practical mismatch measure can be constructed  taking the mean of the eigenvalues of $g_{ij}(\mathcal{A},\Vec{\theta}_t)$ defining an averaged metric as in \cite{normfisher}, $\overline{g}_{ij}(\Vec{\theta}_t)$.

Considering long-duration observation times ($T_{\text{obs}}$ of a few days), the metric $\overline{g}_{ij}(\mathbf{\theta}_t)$ can be approximated by the so-called "phase metric" \cite{metricF}, which neglects amplitude modulation but retains detector-specific phase modulation. In recent decades, several studies have explored the definition and computation of the phase metric. All the results from these studies can be readily applied to the 5-vector, as the phase metric depends solely on the signal phase modulation.

\subsection{Sensitivity estimation}
\begin{figure}[t]
\centering
        \includegraphics[scale=0.7]{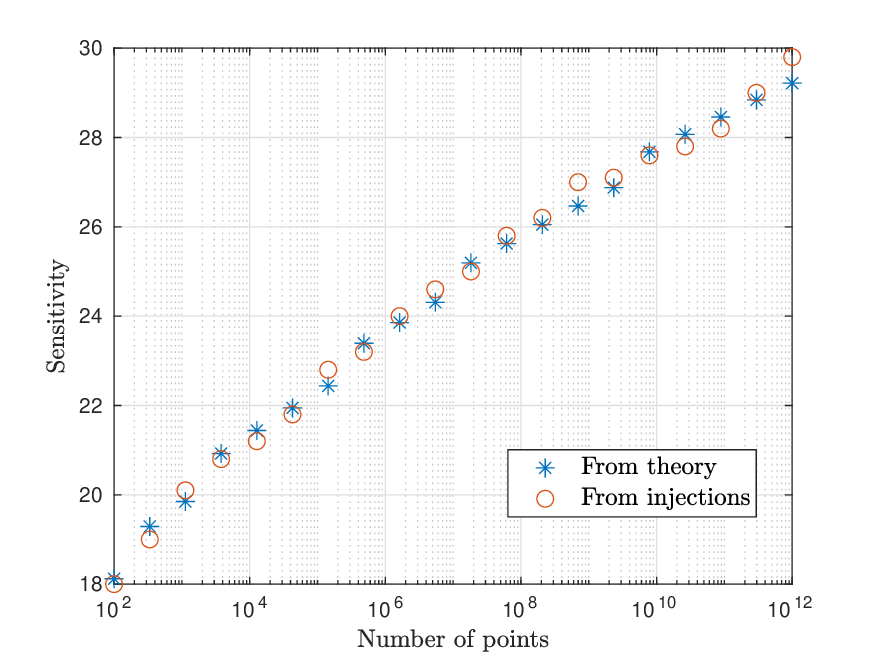}
    \caption{Sensitivity of a narrowband search using the 5n-vector method. The Figure shows (blue stars) the factor $C$ in \eqref{hmin} as a function of the number of points $N$ explored in the parameter space from the theoretical computation in \eqref{hmin_narrow}. The red circles corresponds to the results in Figure 5 of \cite{2014}, inferred from a software injection campaign.}
    \vspace{-10pt}
    \label{narrowsens}
\end{figure}
The sensitivity of a specific search is generally defined as the minimum detectable amplitude:
\begin{equation}\label{hmin}
    h_{\text{min}} \approx C \sqrt{\frac{S_h(f)}{T}}
\end{equation}
where $S_h(f)$ is the power spectral density at frequency $f$, and $T$ the effective observation time. The efficiency factor $C$ depends on the considered search. For example, for a targeted search $C\approx 11$, while for a narrowband search \cite{2014} considering a grid in the frequency-frequency derivative space, $C$ is a function of the number $N$ of points explored in the parameter space (see for example, Figure 5 in \cite{2014}).  In this Section, we show that the factor $C$ can be computed analytically from the theoretical distributions in the assumption of stationary Gaussian noise avoiding Monte-Carlo injections studies. 

The minimum detectable amplitude corresponds to the value of the amplitude that entails a desired value of the detection probability for a fixed false-alarm probability. As shown, the signal distribution is ruled by the non-centrality parameter $\rho^2$, which is proportional to the squared amplitude $H_0^2$. To estimate $h_{\text{min}}$, we can invert the relation in \eqref{lambdanew} fixing the value of $\rho^{95\%,1\%}$ that entails a detection probability of $95\%$ for a false alarm of $1\%$. For single detector:
\begin{equation}
    h_{\text{min}}\approx 1.32 \sqrt{\frac{\rho^{95\%,1\%}}{|\textbf{A}^+|^2 + |\textbf{A}^\times|^2}}\sqrt{\frac{S_h}{T}}
\end{equation}
taking the average over the polarization parameters ($-\pi/2\leq \psi\leq\pi/2$, $-1\leq \cos \iota \leq1$ uniformaly distributed). The factor $1.32$ takes into account the conversion factor between $H_0$ and the standard amplitude $h_0$ \cite{targO2}.  Since the statistic distributions are fixed for any pulsars and any detectors, $\rho^{95\%,1\%} \approx 24$\footnote{The value $\rho^{95\%,1\%} \approx 24$ is inferred fixing the value of the statistic that entails a false alarm of $1\%$ from the noise distribution and then, using this fixed value to select the non-centrality parameter that entails a detection probability of $95\%$ from the signal distribution.} for right ascension and declination that are uniformly distributed, and averaging the theoretical expressions (Eq. 17-18 in \cite{2010}) for $\textbf{A}^{+/\times}$ entails that $|\textbf{A}^+|^2 + |\textbf{A}^\times|^2\approx 0.4$. It follows that:
\begin{equation}
    C \approx 1.32 \sqrt{\frac{\rho^{95\%,1\%}}{0.4}} \approx 10.3\,.
\end{equation}
If we are exploring a parameter space using a template grid with a $N$ points, we need to take into account the trial factors that decrease the search sensitivity and the false alarm, i.e we have to consider $\rho^{95\%,1\%/N}$ (see Figure \ref{narrowsens}, for the relation with $N$). If we are considering a narrowband/directed search for a specific target, we can avoid the average over the sky position and consider the actual value of $|\textbf{A}^+|^2 + |\textbf{A}^\times|^2$.

In the multidetector case, averaging over the sky positions, we have:
\begin{equation}\label{hmin_narrow}
       h_{\text{min}}\approx 1.32 \sqrt{\frac{\rho^{95\%,1\%/N}}{0.4}}\sqrt{\left(\sum\limits_{i=1}^n \frac{T_i}{S_i}\right)^{-1}} \,.
\end{equation}

In a semicoherent search \cite{semicoh}, the observation time is generally divided in $N$ chunks $T_{\text{coh}}$-long (with $T_{\text{coh}}$ a multiple of the sidereal day) and individually analyzed. Then, the detection statistic is defined as the sum of the statistics computed for each chunk. This type of search allows to typically perform more robust analysis since there is no phase continuity between the segments.  To asses the sensitivity, we need to consider the signal distribution of the statistics sum that is a 4N-D $\chi^2$ distribution with non centrality parameter:
\begin{equation}
    \lambda_{\text{SC}} = 2 H_0^2 T_{\text{coh}}\left(|H_+|^2  |\textbf{A}^+|^2 + |H_\times|^2 |\textbf{A}^\times|^2  \right) \sum\limits_{i=1}^N \frac{1}{S_i}
\end{equation} considering, for simplicity, the single detector case. It follows that:
\begin{equation}
   h_{\text{min}}\approx 1.32 \sqrt{\frac{\lambda_{\text{SC}}^{95\%,1\%}}{0.4}}\sqrt{\left(T_{\text{coh}}\sum\limits_{i=1}^N \frac{1}{S_i}\right)^{-1}} \,,
\end{equation}
noting that the $\lambda_{\text{SC}}^{95\%,1\%}$ is fixed considering the distribution of the statistics sum, and $S_i$ is the detector sensitivity in the i-th data chunk.

\section{Conclusion}
In this paper, we derived the 5-vector method and its related detection statistic using a maximum likelihood approach. While the relationship between the 5-vector statistic and the $\mathcal{F}$-statistic was intuitively recognized in \cite{2010} for a single detector, we extended this to multiple detectors, highlighting significant differences from the classic multidetector extension in \cite{2014}. The maximum likelihood approach provides different, statistically efficient estimators of the polarization amplitudes by accounting for the varying sensitivity and observation time of each detector. We demonstrated that the 5-vector statistic derived from the maximum likelihood is statistically equivalent to the $\mathcal{F}$-statistic, as they share the same distributions.

The definition of the data 5-vector introduced in \eqref{5vecnew} from the maximum likelihood estimation adds a factor that is the inverse of the observation time with respect to the classic definition in \cite{2010}. For the multidetector case, the resulting 5n-vector definition in \eqref{5nvec_ML} combines the 5-vector from each detector with weights that are the square root of the ratio between the corresponding observation time and sensitivity. This definition allows to consider multiple detector also if the detectors sensitivity and/or observation time vary significantly between the detectors. 

As a toy case, approximating the current observing run O4 for the LIGO-Virgo detectors, we can assume equal sensitivity $S_h$ and observation time $T$ for the two LIGO. Since Virgo joined the O4 run in the second part with higher sensitivity, we can assume $A\cdot S_h$ and $T/2$ for Virgo. The factor $A$ depends on the frequency and generally, it should be between $2\leq A \leq5$. Using the sensitivity estimation in \eqref{hmin_narrow}, considering only the two LIGO detectors we have
\begin{equation}
    h_{\text{min}}^{(2)}\approx C \sqrt{\frac{S_h}{T}}\frac{1}{\sqrt{2}} \,,
\end{equation}while including Virgo,
\begin{equation}
    h_{\text{min}}^{(3)}\approx C \sqrt{\frac{S_h}{T}}\sqrt{\frac{2A}{4A+1}} \,.
\end{equation}
It follows that $h_{\text{min}}^{(3)}\approx h_{\text{min}}^{(2)}\sqrt{4A/(4A+1)}$; i.e there is an improvement between $6\%$ for $A=2$ and $2.5\%$ for $A=5$ including Virgo data with respect to consider only the two LIGO detectors for the O4 run.

For future CW searches using the 5-vector formalism, we recommend using the 5n-vector from the maximum likelihood approach for multidetector analysis. This approach offers a multidetector extension that always improves the detection sensitivity, efficient estimators for the polarization amplitudes, analytical theoretical distributions for the detection statistic, and a solid framework for constructing a template grid. Additionally, we provided an analytical computation of the sensitivity for different searches, estimating the minimum detectable amplitude. These theoretical estimations will simplify and significantly reduce the computational cost of search sensitivity estimation, bypassing extensive Monte-Carlo analyses.

\singlespacing
\printbibliography 

\end{document}